\documentclass[lettersize,journal]{IEEEtran}
\usepackage{amsmath,amsfonts}
\usepackage{algorithmic}
\usepackage{algorithm}
\usepackage{array}
\usepackage[caption=false,font=normalsize,labelfont=sf,textfont=sf]{subfig}
\usepackage{textcomp}
\usepackage{stfloats}
\usepackage{url}
\usepackage{verbatim}
\usepackage{graphicx}
\usepackage{cite}
\usepackage{xcolor}
\usepackage{float}
\usepackage{hyperref}					
\hypersetup{colorlinks,
	linkcolor=blue,%
	anchorcolor=blue,
    citecolor=blue}

\begin{document}

\title{\huge Interference-Resilient OFDM Waveform Design with Subcarrier Interval Constraint for ISAC Systems}

\author{Qinghui Lu,~Zhen Du,~\IEEEmembership{Member,~IEEE,}~and~Zenghui Zhang,~\IEEEmembership{Senior Member,~IEEE}
\thanks{This work was supported in part by the National Natural Science Foundation of China under Grants 62271311 and 62301264, and in part by the Natural Science Foundation of Jiangsu Province under Grant BK20230416.}
\thanks{Qinghui Lu and Zenghui Zhang are with the Shanghai Key Laboratory of Intelligent Sensing and Recognition, Shanghai Jiao Tong University, Shanghai 200240, China (e-mail: zenghui.zhang@sjtu.edu.cn).}
\thanks{Zhen Du is with the School of Electronic and Information Engineering, Nanjing University of Information Science and Technology, Nanjing 210044, China.}
}

\markboth{Journal of \LaTeX\ Class Files,~Vol.~xx, No.~xx, December~2023}%
{Shell \MakeLowercase{\textit{et al.}}: A Sample Article Using IEEEtran.cls for IEEE Journals}


\maketitle

\begin{abstract}
Conventional orthogonal frequency division multiplexing (OFDM) waveform design in integrated sensing and communications (ISAC) systems usually selects the channels with high-frequency responses to transmit communication data, which does not fully consider the possible interference in the environment. To mitigate these adverse effects, we propose an optimization model by weighting between peak sidelobe level and communication data rate, with power and communication subcarrier interval constraints. To tackle the resultant nonconvex problem, an iterative adaptive cyclic minimization (ACM) algorithm is developed, where an adaptive iterative factor is introduced to improve convergence. Subsequently, the least squares algorithm is used to reduce the coefficient of variation of envelopes by further optimizing the phase of the OFDM waveform. Finally, the numerical simulations are provided to demonstrate the interference-resilient ability of the proposed OFDM strategy and the robustness of the ACM algorithm.
\end{abstract}


\section{Introduction}\label{In}
\IEEEPARstart{I}{ntegrated} sensing and communications (ISAC) as an enabler to synergistically design sensing and communications (S\&C) functionalities, can facilitate the utilization efficiency of both hardware and wireless resources \cite{ISAC1}, which has been envisioned as a promising technology for numerous emerging applications in 6G networks, such as intelligent transportation, activity recognition, smart city, etc \cite{ea2,fe,ISAC4}.

To attain excellent S\&C performance, waveform design approaches are desired to facilitate communication data rate (CDR) and sensing capabilities such as target detection, estimation, and tracking. Consequently, one of the best waveform candidates is orthogonal frequency division multiplexing (OFDM), owing to its superiority of simple discrete Fourier transform (DFT) structure, large bandwidth enabling high CDR and range resolution, and frequency diversity, etc\cite{ofdm1,ofdm2,ofdm4,ofdm5,papr1}.
For instance, the authors in \cite{papr1} designed a peak-to-average power ratio (PAPR) reduction scheme under the principle of uniform power allocation, which only optimizes the integrated sidelobe level of the autocorrelation function, while resulting in limited communication performance. To this end, in \cite{ofdm1}, the power minimization-based joint subcarrier assignment and power allocation (SAPA) model is formulated while guaranteeing the specified S\&C constraints. However, the subcarrier assignment strategy aims to transmit data through the communication channels with a high signal-to-noise ratio (SNR), while its performance may be impacted by potential interference in practical scenarios.
An integrated OFDM waveform method to reduce the peak sidelobe level (PSL) while meeting the CDR requirement is considered in \cite{ofdm2}. Nevertheless, the proposed algorithm in \cite{ofdm2} is a heuristic approach whose results are susceptible to the initial feasible points (IFPs), so this algorithm is not robust.

From the aforementioned discussions, OFDM waveform design methods applied to ISAC systems are in the absence of a more comprehensive model and a more robust algorithm. In this letter, we present an OFDM waveform optimization strategy with two steps. Firstly, a joint SAPA method for optimizing autocorrelation PSL and CDR under the constraints of power and communication subcarrier interval is established. To solve this nonconvex problem, a modified adaptive cyclic minimization (ACM) algorithm is proposed, and an iteration factor is introduced to release the effect caused by the IFPs. Then, taking the coefficient of variation of envelopes (CVE) \cite{papr2} as the objective function to minimize, we optimize the remaining phase, except for which is occupied by the communication symbols, in order to mitigate the envelope fluctuation. Simulation results demonstrate the effectiveness of the proposed method.

The rest of this letter is organized as follows. In section \ref{OSMM}, we introduce the OFDM signal model and S\&C metrics. We propose a joint SAPA method in section \ref{JSPAM}. In section \ref{CDM}, we model the CVE reduction problem and develop an iterative algorithm to solve it. In section \ref{SR}, we evaluate the performance of the proposed method by numerical simulations. This letter is finally concluded in section \ref{Conclu}.

\textit{Notation}: The transpose and Hermitian operators are denoted by ${\left(  \cdot  \right)^T}$ and ${\left(  \cdot  \right)^H}$, respectively. The modulus of a complex number is denoted by $\left|  \cdot  \right|$, the Euclidean norm is denoted by $\left\|  \cdot  \right\|$, and ${\left\|  \cdot  \right\|_1}$ means the $l_1$-norm. $\mathbb{R}$ and $\mathbb{C}$ represent the real and complex set, respectively. $\odot$ denotes Hadamard product. $\mathbb{E}\left[  \cdot  \right]$ is mathematical expectation. ${\rm diag}\left[  \cdot  \right]$ represents the diagonal matrix. 

\section{OFDM Signal Model and S\&C Metrics}\label{OSMM}
In this section, we first discuss the system model and then introduce a generic OFDM waveform structure where the communication frequency bins are allocated over a large contiguous radar band. Finally, we introduce the S\&C metrics.

\begin{figure}[t]
	\centerline{\includegraphics[width=3in]{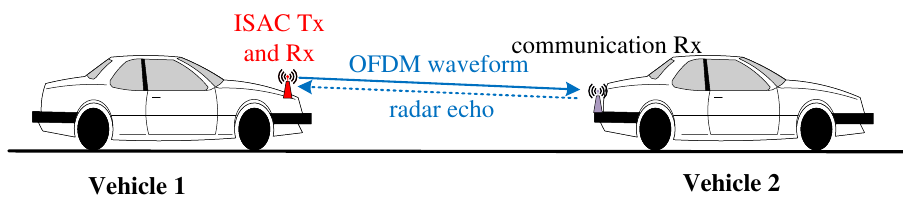}}\vspace{-0.3cm}
	\caption{A representative vehicular framework in ISAC systems.}
	\label{fig1}
\end{figure}
Consider the system model in a typical vehicular scenario presented in Fig. 1. Vehicle 1 is equipped with an ISAC transceiver that radiates an integrated OFDM waveform for radar detection and communication transmission. Specifically, Vehicle 1 can send communication symbols to the communication receiver of Vehicle 2 and estimate target information such as the range and speed of Vehicle 2.

\subsection{Signal Model}\label{ISM}
The integrated OFDM waveform, yielding $N$ subbands in frequency domain, can be defined as \cite{papr1}
\begin{equation}
{\bf{s}} = {{\bf{F}}^H}{\bf{x}} = {{\bf{F}}^H}\left[ {{\bf{Uc}} + \left( {{\bf{I}} - {\bf{U}}} \right){\bf{r}}} \right],
\label{OFDMtimedomain}
\end{equation}
where ${\bf{F}} \in {\mathbb{C}^{N \times N}}$ represents the DFT matrix, ${{\bf{F}}_{k,p}} = {e^{ - j\frac{{2\pi }}{N}kp}},\left( {k,p = 0, \cdots ,N-1} \right)$. And ${\bf{x}}= {{\bf{Uc}} + \left( {{\bf{I}} - {\bf{U}}} \right){\bf{r}}} $ stands for sensing symbols in frequency domain, where the phase part of ${\bf{c}} \in {\mathbb{C}^{N \times 1}}$ bears binary communication data modulated by a phase shift keying (PSK) modulator, and the phase part of ${\bf{r}} \in {\mathbb{C}^{N \times 1}}$ is reserved to mitigate envelope fluctuations. ${\bf{U}} = {\mathop{\rm diag}\nolimits}\left[ {\bf{u}} \right] $ selects the subcarriers for communication, in which the selection variable ${\bf{u}} \in {\mathbb{R}^{N \times 1}}$ is binary with entry one denoting the corresponding subcarrier selected for communication and entry zero discarded.

We also apply the ISAC signal processing structure referring to \cite{papr1}. Particularly, the sensing echo of ${\bf{s}}$ is received and processed for target detection. And for communication reception, the symbols corresponding to the phase part of $\bf c$ should be extracted for PSK demodulation to obtain binary data.

\subsection{Sensing Metric: Autocorrelation PSL}\label{RM}
To suppress interference and improve target detection capability, low autocorrelation PSL property is highly required, which is expressed as \cite{ofdm2}
\begin{equation}
{\mathop{\rm PSL}\nolimits}  = \mathop {\max }\limits_{k \in {\bf{\Theta }}}\left| {{R_{\bf{s}}}\left[ k \right]} \right|= \mathop {\max }\limits_{k \in {\bf{\Theta }}} \left| {\sum\nolimits_{n = 0}^{N - 1} {{{\left| {{{{x}}_n}} \right|}^2}{e^{{{j\pi nk} \mathord{\left/
								{\vphantom {{j\pi nk} K}} \right.
								\kern-\nulldelimiterspace} K}}}} } \right|,
	\label{Rpsl}
\end{equation}
where ${R_{\bf{s}}}\left[ k \right],k \in \left[ { - K + 1,K - 1} \right]$ denotes the $\left( {2K - 1} \right)$ autocorrelation sampling points, ${\bf{\Theta }} = \left[ { - K + 1,\Upsilon } \right) \cup \left( {\Upsilon ,K - 1} \right]$ is the sidelobe region with $\Upsilon$ being the mainlobe boundary.

\subsection{Communication Metric 1: CDR}\label{CM1}
In the frequency selective fading channel, the CDR is regarded as a significant communication metric, which can be optimized by selecting appropriate subcarriers and allocating corresponding transmit power. The definition of CDR is   
\begin{equation}
{\mathop{\rm CDR}\nolimits}  = \sum\nolimits_{n = 0}^{N - 1} {\log_2 \left[ {1 + {{{u_n}{{\left| {{c_n}} \right|}^2}{{\left| {{h_n}} \right|}^2}} \mathord{\left/
				{\vphantom {{{u_n}{{\left| {{c_n}} \right|}^2}{{\left| {{h_n}} \right|}^2}} {\sigma _c^2}}} \right.
				\kern-\nulldelimiterspace} {\sigma _c^2}}} \right]}, 
\label{Cdef}
\end{equation}
where $h_n$ means the frequency response of the $n$-th subcarrier, and $\sigma _c^2$ denotes the noise power in communication channel.

\subsection{Communication Metric 2: CVE}\label{CM2}
Generally speaking, the amplitude variation of the OFDM signal fluctuates wildly, which will lead to signal distortion and increase the bit error rate (BER). The authors in \cite{papr1} improve this shortcoming by reducing the PAPR of the OFDM waveform, which is defined as the ratio between the maximum power and its average power. It is worth noting that this criterion only seeks to decrease the peak values in $\bf s$. Herein, we adopt the CVE as the other communication metric, which takes into account both peak and valley, defined as \cite{papr2}
\begin{equation}
{\rm{CVE}} = {{\mathbb{E}\left[ {{{\left( {\left| {{s_n}} \right| - \mathbb{E}\left[ {\left| {{s_n}} \right|} \right]} \right)}^2}} \right]} \mathord{\left/
		{\vphantom {{\mathbb{E}\left[ {{{\left( {\left| {{s_n}} \right| - \mathbb{E}\left[ {\left| {{s_n}} \right|} \right]} \right)}^2}} \right]} {{{\left( {\mathbb{E}\left[ {\left| {{s_n}} \right|} \right]} \right)}^2}}}} \right.
		\kern-\nulldelimiterspace} {{{\left( {\mathbb{E}\left[ {\left| {{s_n}} \right|} \right]} \right)}^2}}}.
	\label{cve}
\end{equation}

Notably, the autocorrelation PSL and CDR are only related to the transmit power ${p_n} = {\left| {{x_n}} \right|^2}$ and communication power ${p_{c,n}} = {\left| {{c_n}} \right|^2}$. This inspires us to devise a joint SAPA method to improve the ISAC performance, and reduce the CVE by further optimizing the phase of $\bf r$.

\section{Joint Subcarrier Assignment and Power Allocation Strategy}\label{JSPAM}
In this section, we investigate a more comprehensive joint design optimization problem for ISAC systems, and propose a new ACM algorithm to solve it.

\subsection{Problem Formulation}\label{PF1}
In this subsection, the transmit power ${p_n}$ and communication subcarrier indicator $u_n$ are optimized. Specifically, we build up an optimization problem of joint autocorrelation PSL and CDR, which is formulated as
\begin{align}\label{prob1}
	\mathop {\min }\limits_{{p_n},{u_n}} \ & \rho \frac{{\mathop {\max }\limits_{k \in {\bf{\Theta }}} \!\left| {\sum\limits_{n = 0}^{N - 1}\! {{p_n}{e^{{{j\pi nk} \mathord{\left/
									{\vphantom {{j\pi nk} K}} \right.
									\kern-\nulldelimiterspace} K}}}} } \right|}}{{{R_{\max }}}} \!- \!\left( {1 \!-\! \!\rho } \right)\!\frac{{\sum\limits_{n = 0}^{N - 1}\! {{{\log }_2}\!\left[ {1 + \frac{{{u_n}{p_n}\!{{\left| {{h_n}} \right|}^2}}}{{\sigma _c^2}}} \right]} }}{{{C_{\max }}}} \nonumber \\
	{\rm s.t.} \ & {{\mathop{\rm C}\nolimits} _1}:\sum\nolimits_{n = 0}^{N - 1} {{p_n}}  = {P_{total}},\nonumber \\
	& {{\mathop{\rm C}\nolimits} _2}:\sum\nolimits_{n = 0}^{N - 1} {{u_n}{p_n}}  \le {P_c},\nonumber \\
	& {{\mathop{\rm C}\nolimits} _3}:{p_n} \ge 0,\forall n \in N,\\
	& {{\mathop{\rm C}\nolimits} _4}:{u_n} \in \left\{ {0,1} \right\},\nonumber \\
	& {{\mathop{\rm C}\nolimits} _5}:\sum\nolimits_{n = 0}^{N - 1} {{u_n}}  = {N_r},\nonumber \\
	& {{\mathop{\rm C}\nolimits} _6}:\!{u_{i + L}} \!+ \! \cdots  \!+ {u_{i + 1}} \!+\! {u_i} \!\le\! 1,i\! = \!0,\! \cdots\! ,N\! -\!\! L\!-\!1.\nonumber 
\end{align}
where $R_{\max }$ and $C_{\max }$ denote the maximum autocorrelation sidelobe level without optimizing, and the maximum CDR, respectively. $\rho  \in \left[ {0,1} \right]$ is a weighted coefficient striking a trade-off between autocorrelation PSL and CDR, thereby balancing S\&C performance. $P_{total}$ denotes the total transmit power of all subcarriers. $P_c$ represents the threshold of maximum communication power. The constraints ${{\mathop{\rm C}\nolimits} _4}$ and ${{\mathop{\rm C}\nolimits} _5}$ indicate that only $N_r$ subcarriers are allocated for communication purpose. Notably, the minimum interval constraint ${{\mathop{\rm C}\nolimits} _6}$ implies that the index interval between two adjacent communication subcarriers is no less than $\left( {L + 1} \right)$. Innovatively, we exploit the constraint ${{\mathop{\rm C}\nolimits} _6}$ to reduce the probability of interference, mainly concentrating on the case that non-cooperators impose interference based on channel characteristics.

\subsection{Adaptive Cyclic Minimization Algorithm}\label{A1}
Due to the fact that variable $u_n$ is binary, the resulting joint optimization model is a mixed-integer nonconvex problem. Since the two optimization variables $u_n$ and $p_n$ can be solved separately \cite{suan} in each iteration, a tailored ACM algorithm is utilized to tackle the problem \eqref{prob1} by solving the subproblems of $u_n$ and $p_n$ in sequence.

To be more specific, we firstly introduce the auxiliary variable $\eta $ and then rewrite problem \eqref{prob1} in the following form:
\begin{equation}
\begin{aligned}
	\mathop {\min }\limits_{{p_n},{u_n},\eta } \ & {\rm{ }}\rho \frac{\eta }{{{R_{\max }}}} - \left( {1 - \rho } \right)\frac{{\sum\nolimits_{n = 0}^{N - 1} {{{\log }_2}\left[ {1 + \frac{{{{\left| {{h_n}} \right|}^2}{u_n}{p_n}}}{{\sigma _c^2}}} \right]} }}{{{C_{\max }}}}\\
	{\rm s.t.}\ & {{\mathop{\rm C}\nolimits} _1},{{\mathop{\rm C}\nolimits} _2},{{\mathop{\rm C}\nolimits} _3},{{\mathop{\rm C}\nolimits} _4},{{\mathop{\rm C}\nolimits} _5},{{\mathop{\rm C}\nolimits} _6},\\
	& {{\mathop{\rm C}\nolimits} _7}:\left| {\sum\nolimits_{n = 0}^{N - 1} {{p_n}{e^{{{j\pi nk} \mathord{\left/
							{\vphantom {{j\pi nk} K}} \right.
							\kern-\nulldelimiterspace} K}}}} } \right| \le \eta ,k \in \bf{\Theta } .\\
\end{aligned}
\label{prob11}
\end{equation}

Due to the constraints ${{\mathop{\rm C}\nolimits} _4}$ and ${{\mathop{\rm C}\nolimits} _5}$, the suboptimal value of $u_n$ can be obtained by utilizing either exhaustive search or heuristic search method \cite{ofdm2}. Subsequently, for a specified value of $u_n$, the transmit power $p_n$ can be determined. Assuming that $p_n^{\left( t \right)},u_n^{\left( t \right)},{\eta ^{\left( t \right)}}$ are obtained at the $t$th iteration, optimization variables at the $\left( {t + 1} \right)$th iteration can be updated via the following two steps.

\subsubsection{Step 1 - Updating $u_n^{\left( t+1 \right)}$}\label{s1}
Ignoring irrelevant terms, the subproblem with respect to $u_n$ can be simplified as
\begin{equation}
\begin{aligned}
    \mathop {\min }\limits_{{u_n}} \ & {\rm{ }} - \sum\nolimits_{n = 0}^{N - 1} {{{\log }_2}\left[ {1 + \frac{{{{\left| {{h_n}} \right|}^2}{u_n}p_n^{\left( t \right)}}}{{\sigma _c^2}}} \right]} \\
	{\rm s.t.}\ & {{\mathop{\rm C}\nolimits} _2},{{\mathop{\rm C}\nolimits} _5},{{\mathop{\rm C}\nolimits} _6},\\
	& {{\mathop{ \rm C}\nolimits} _4}:{u_n} \in \left\{ {0,1} \right\}.
\end{aligned}\label{sub11}
\end{equation}

The nonconvex constraint ${{\mathop{ \rm C}\nolimits} _4}$ is equivalent to \cite{apro}
\begin{equation}
\begin{aligned}
	\mathop {\min }\limits_{\bf{u}} \ {{\bf{u}}^T}\left( {1 - {\bf{u}}} \right) \ 
	{\rm s.t.}\ 0 \le {u_n} \le 1,\forall n.
\end{aligned}	
\label{sub12}
\end{equation}
However, the objective function in \eqref{sub12} is concave and difficult to solve. Resorting to the first-order Taylor expansion around ${{\bf{u}}^{\left( {t + 1,m - 1} \right)}}$ (the result of ${{\bf{u}}^{\left( {t + 1} \right)}}$ at the $\left( {m - 1} \right)$th iteration), the problem \eqref{sub11} can be approximated as
\begin{align}\label{sub13}
	\mathop {\min }\limits_{{u_n}} \ & - \sum\nolimits_{n = 0}^{N - 1} {{{\log }_2}\left[ {1 + \frac{{{{\left| {{h_n}} \right|}^2}{u_n}p_n^{\left( t \right)}}}{{\sigma _c^2}}} \right]} \nonumber \\ & + \lambda \left[ {{{\bf{u}}^T}\!\left( {1\! - \!2{{\bf{u}}^{\left( {t + 1,m - 1} \right)}}} \right) \!+\! {{\bf{u}}^{{{\left( {t + 1,m - 1} \right)}^T}}}\!{{\bf{u}}^{\left( {t + 1,m - 1} \right)}}} \right] \nonumber \\
	{\rm s.t.}\ & {{\mathop{\rm C}\nolimits} _2},{{\mathop{\rm C}\nolimits} _5},{{\mathop{\rm C}\nolimits} _6}, \nonumber \\
& {{\mathop{ \rm\bar C}\nolimits} _4}:0 \le {u_n} \le 1,\forall n.
\end{align}
where $\lambda$ represents a weight factor. This convex problem can be solved iteratively via interior point method (IPM), and can be implemented by the CVX toolbox \cite{CVX1}.

Since the IFP of $u_n$ and the value of $\lambda$ significantly affect the algorithm convergence, we ameliorate this shortcoming by adaptively updating $\lambda$ according to the following iteration formula\cite{adp}
\begin{equation} 
{\lambda ^{\left( t + 1,m \right)}} \!= \!\left\{ \!\begin{array}{l}
	\!{\lambda ^{\left( {t + 1,m - 1} \right)}},{\alpha ^{\left( {t + 1,m - 1} \right)}} \le {\xi _1}{\alpha ^{\left( {t + 1,m - 2} \right)}}\\
	\!{\xi _2}{\lambda ^{\left( {t + 1,m - 1} \right)}},\ {\rm{otherwise}}
\end{array} \right.
	\label{sub14}
\end{equation}
where ${\xi _1} < 1$, ${\alpha ^{\left( {t + 1,m} \right)}} = {{\bf{u}}^{{{\left( {t + 1,m} \right)}^T}}}\left( {1 - 2{{\bf{u}}^{\left( {t + 1,m - 1} \right)}}} \right) + {{\bf{u}}^{{{\left( {t + 1,m - 1} \right)}^T}}}{{\bf{u}}^{\left( {t + 1,m - 1} \right)}}$ and ${\xi _2} > 1$. As a consequence, ${\alpha ^{\left( {t + 1,m} \right)}}$ tends to 0 during iterations and parameter tuning can be avoided.

In addition, the exit condition is defined as ${\alpha ^{\left( {t + 1,m} \right)}} \le {\varepsilon _u}$.

\subsubsection{Step 2 - Updating $p_n^{\left( {t + 1} \right)},{\eta ^{\left( {t + 1} \right)}}$}\label{s2}
With a determined ${\bf u}^{\left( {t + 1} \right)}$, the optimization problem with respect to $p_n$ and $\eta$ can be expressed as
\begin{equation} 
\begin{aligned}
    \mathop {\min }\limits_{{p_n},\eta } \ & \rho \frac{\eta }{{{R_{\max }}}} - \left( {1 - \rho } \right)\frac{{\sum\nolimits_{n = 0}^{N - 1} {{{\log }_2}\left[ {1 + \frac{{{{\left| {{h_n}} \right|}^2}u_n^{\left( {t + 1} \right)}{p_n}}}{{\sigma _c^2}}} \right]} }}{{{C_{\max }}}}\\
	{\rm s.t.} \ & {{\mathop{\rm C}\nolimits} _1},{{\mathop{\rm C}\nolimits} _3},{{\mathop{\rm C}\nolimits} _7},\\ 
    & {{\mathop{ \rm\bar C}\nolimits} _2}:\sum\nolimits_{n = 0}^{N - 1} {u_n^{\left( {t + 1} \right)}{p_n}}  \le {P_c}.
\end{aligned}
	\label{sub21}
\end{equation}
Evidently, it is a convex optimization problem that can be solved using the CVX toolbox \cite{CVX1}.

Finally, we repeat step 1 and step 2 until the maximum number of iterations $t_{\max}$ is reached, or $\Delta r_c^{\left( t \right)} \le \varepsilon _c$ and $\Delta r_a^{\left( t \right)} \le \varepsilon _a$  are satisfied at the same time, where $\varepsilon _c$ and $\varepsilon _a$ are the maximum tolerance errors of CDR and PSL, respectively. Notably, the residuals $\Delta r_c$ and $\Delta r_a$ are defined as
\begin{equation} 
\Delta r_c^{\left( t+1 \right)} = \left| {{{\left( {Obj_c^{\left( t+1 \right)} - Obj_c^{\left( {t} \right)}} \right)} \mathord{\left/
			{\vphantom {{\left( {Obj_c^{\left( t+1 \right)} - Obj_c^{\left( {t} \right)}} \right)} {Obj_c^{\left( t \right)}}}} \right.
			\kern-\nulldelimiterspace} {Obj_c^{\left( t \right)}}}} \right|,
	\label{Dc1}
\end{equation}
\begin{equation} 
Obj_c^{\left( t \right)} = \sum\limits_{n = 0}^{N - 1} {{{\log }_2}\left[ {1 + {{{{\left| {{h_n}} \right|}^2}u_n^{\left( t \right)}p_n^{\left( t \right)}} \mathord{\left/
				{\vphantom {{{{\left| {{h_n}} \right|}^2}u_n^{\left( t \right)}p_n^{\left( t \right)}} {\sigma _c^2}}} \right.
				\kern-\nulldelimiterspace} {\sigma _c^2}}} \right]}, 
	\label{Dc2}
\end{equation}
\begin{equation} 
\Delta r_a^{\left( t+1 \right)} = \left| {{{\left( {Obj_a^{\left( t+1 \right)} - Obj_a^{\left( {t } \right)}} \right)} \mathord{\left/
			{\vphantom {{\left( {Obj_a^{\left( t+1 \right)} - Obj_a^{\left( {t} \right)}} \right)} {Obj_a^{\left( t \right)}}}} \right.
			\kern-\nulldelimiterspace} {Obj_a^{\left( t \right)}}}} \right|,
	\label{Da1}
\end{equation}
\begin{equation} 
	Obj_a^{\left( t \right)} = \mathop {\max }\limits_{k \in \bf{\Theta } } \left| {\sum\limits_{n = 0}^{N - 1} {{p_n^{\left( t \right)}}{e^{{{j\pi nk} \mathord{\left/
							{\vphantom {{j\pi nk} K}} \right.
							\kern-\nulldelimiterspace} K}}}} } \right|.
	\label{Da2}
\end{equation}

For the sake of completeness, the main steps of the proposed ACM algorithm are summarized in Algorithm \ref{alg:1}.
\begin{algorithm}
	\renewcommand{\algorithmicrequire}{\textbf{Input:}}
	\renewcommand{\algorithmicensure}{\textbf{Output:}}
	\caption{ACM Algorithm for Solving Problem \eqref{prob1}}
	\label{alg:1}
	\begin{algorithmic}[1] 
		\REQUIRE $\rho$, $h_n$, ${\sigma _c^2}$, $N$, $P_{total}$, $P_c$, $N_r$, $L$, ${\lambda ^{\left(0 \right)}}$, ${\xi _1}$, ${\xi _2}$, $\varepsilon _c$, $\varepsilon _a$;
		\ENSURE ${\bf{u}},{\bf{p}}$;
		\STATE Initialize ${{\bf{u}}^{(0)}}$;
		\FOR {$t = 0, \cdots ,t_{\max}-1$}
		\STATE Step 1: Update $u_n^{\left( t+1 \right)}$ per Eq. \eqref{sub13};\ 
		\STATE Step 2: Update $p_n^{\left( t+1 \right)}$ and $\eta^{\left( t+1 \right)}$ per Eq. \eqref{sub21};\
		\IF {$\Delta r_c^{\left( t+1 \right)} \le \varepsilon _c$ and $\Delta r_a^{\left( t+1 \right)} \le \varepsilon _a$} 
		\STATE Break;\
		\ENDIF
		\ENDFOR
		\STATE Return ${\bf{u}} = {{\bf{u}}^{(t + 1)}}$ and ${\bf{p}} = {{\bf{p}}^{(t + 1)}}$.
	\end{algorithmic}
\end{algorithm}

\section{CVE Design Method}\label{CDM}
In this section, we consider optimizing CVE for ISAC systems. The phase part of $\bf r$ is optimized on the premise that the power of each subcarrier and the modulated communication information are known. 

Omitting the denominator in \eqref{cve}, the CVE-minimization problem can be designed as
\begin{equation}
\begin{array}{l}
	\mathop {\min }\limits_{{{\bf{\Phi }}_r}} {\rm{ }} \ \mathbb{E}\left[ {{{\left( {\left| {{s_n}} \right| - \mathbb{E}\left[ {\left| {{s_n}} \right|} \right]} \right)}^2}} \right]\\
	{\rm s.t.}{\kern 2pt}{\bf{s}} = {{\bf{F}}^H}\!\left[ {{\bf{U}}\!\left( {\sqrt {{{\bf{p}}}} \! \odot\! {e^{j{{\bf{\Phi }}_c}}}} \right) \!\!+\! \left( {{\bf{I}}\! -\! {\bf{U}}} \right)\!\!\left( {\sqrt {{{\bf{p}}}}  \odot {e^{j{{\bf{\Phi }}_r}}}} \right)} \!\right]
\end{array}
\label{prob2}
\end{equation}
where ${{\bf{\Phi }}_r}$ represents the optimizable phase of $\bf r$, and ${e^{j{{\bf{\Phi }}_c}}}$ represents communication symbols drawn from PSK constellation. To proceed, problem \eqref{prob2} can be reformulated as
\begin{equation} 
\begin{aligned}
	\mathop {\min }\limits_{\bf{w}} \ & {\left( {{{\bf{F}}^H}{\bf{w}} + {\bf{v}} - \beta {e^{j{\bf{\Phi }}}}} \right)^H}\left( {{{\bf{F}}^H}{\bf{w}} + {\bf{v}} - \beta {e^{j{\bf{\Phi }}}}} \right)\\
	{\rm s.t.} \ & \beta  = {{{{\left\| {{{\bf{F}}^H}{\bf{w}} + {\bf{v}}} \right\|}_1}} \mathord{\left/
			{\vphantom {{{{\left\| {{{\bf{F}}^H}{\bf{w}} + {\bf{v}}} \right\|}_1}} N}} \right.
			\kern-\nulldelimiterspace} N},\\
	& {\bf{\Phi }}{\rm{ = }}\angle \left( {{{\bf{F}}^H}{\bf{w}} + {\bf{v}}} \right),\\
	& \left| {\bf{w}} \right| = \left( {{\bf{I}} - {\bf{U}}} \right) \odot \sqrt {{{\bf{p}}}} .
\end{aligned}
\label{prob21}
\end{equation}
where $\angle \left(  \cdot  \right)$ denotes the angle of the complex-value, and ${\bf{v}} = {{\bf{F}}^H}{\bf{U}}\left( {\sqrt {{{\bf{p}}}}  \odot {e^{j{{\bf{\Phi }}_c}}}} \right)$ is the known communication part.

To solve \eqref{prob21} at the $\left( {i + 1} \right)$th iteration, \eqref{prob22}-\eqref{prob24} can be iteratively updated via the least squares (LS) algorithm\cite{papr2}:
\begin{equation} 
\beta^{\left( i \right)}  = {{{{\left\| {{{\bf{F}}^H}{\bf{w}}^{\left( i \right)} + {\bf{v}}} \right\|}_1}} \mathord{\left/
		{\vphantom {{{{\left\| {{{\bf{F}}^H}{\bf{w}}^{\left( i \right)} + {\bf{v}}} \right\|}_1}} N}} \right.
		\kern-\nulldelimiterspace} N},
	\label{prob22}
\end{equation}
\begin{equation} 
{{\bf{\Phi }}^{\left( i \right)}}=\angle \left( {{{\bf{F}}^H}{{\bf{w}}^{\left( i \right)}} + {\bf{v}}} \right),
\label{prob23}
\end{equation}
\begin{equation} 
{{\bf{w}}^{\left( {i + 1} \right)}} = \left( {{\bf{I}} - {\bf{U}}} \right) \odot \sqrt {\bf{p}}  \odot \frac{{ - {\bf{F}}\left( {{\bf{c}} - {\beta ^{\left( i \right)}}{e^{j{{\bf{\Phi }}^{\left( i \right)}}}}} \right)}}{{\left| {{\bf{F}}\left( {{\bf{c}} - {\beta ^{\left( i \right)}}{e^{j{{\bf{\Phi }}^{\left( i \right)}}}}} \right)} \right|}}.
\label{prob24}
\end{equation}

\section{Simulation Results}\label{SR}
In this section, we evaluate the proposed method. The parameters are listed as follows. $N = 128$, $N_r = 16$, ${P_{total}} = 256 {\rm{W}}$, ${P_c} = {{{P_{total}}} \mathord{\left/
		{\vphantom {{{P_{total}}} 4}} \right.
		\kern-\nulldelimiterspace} 4}$, $L=5$, $\rho  = 0.5$, ${t_{\max }} = {10^3}$, $\varepsilon _c=\varepsilon _a={10^{ - 4}}$ and ${\bf{\Theta }} = \left[ { - \left( {N - 1} \right):1: - 2,2:1:\left( {N - 1} \right)} \right]$. The Gaussian white noise with power $\sigma _c^2 = 1$ is exploited, and the communication symbols are drawn from the 8-PSK constellation randomly.
	
\subsection{Interference-Resilient Performance Evaluation}\label{MVC}

\begin{figure}[htbp]\vspace{-0.5cm}
	\centerline{\includegraphics[width=3in]{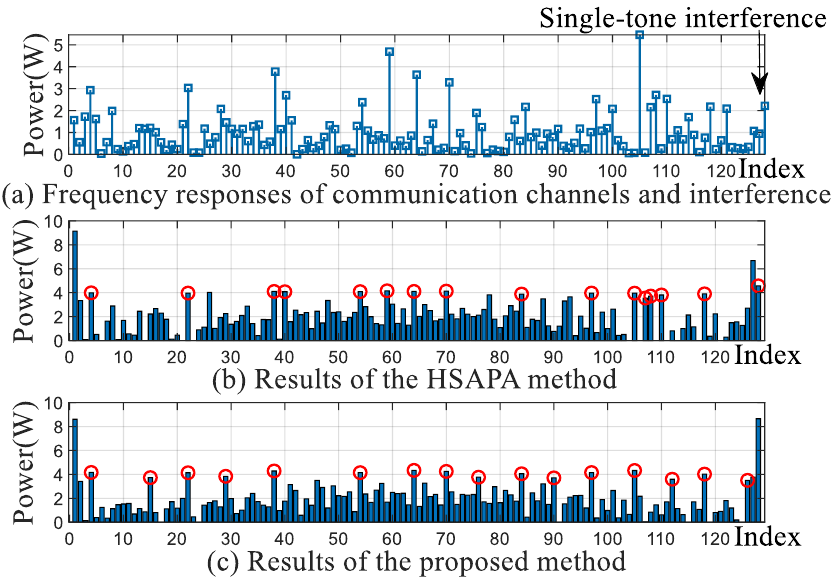}}\vspace{-0.2cm}
	\caption{Comparisons under Rayleigh distribution communication channel: (a) Frequency responses of communication channels and interference, (b) Results of the HSAPA method, (c) Results of the proposed method.}
	\label{Fig1}
\end{figure}

\begin{figure}[htbp]\vspace{-0.5cm}
	\centerline{\includegraphics[width=3in]{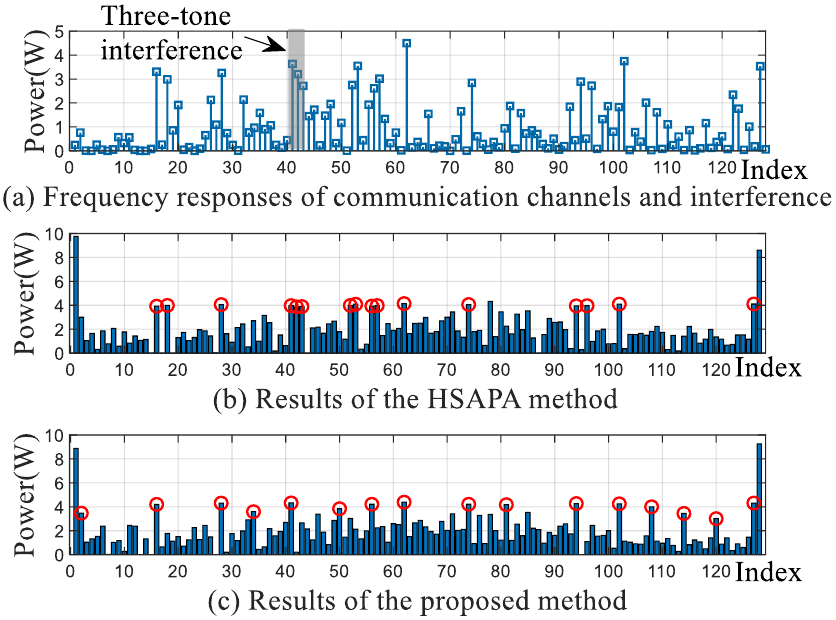}}\vspace{-0.2cm}
	\caption{Comparisons under standard normal distribution: (a) Frequency responses of communication channels and interference, (b) Results of the HSAPA method, (c) Results of the proposed method.}
	\label{Fig2}
\end{figure}

Two types of communication channels, namely the Rayleigh distribution channel and the standard normal distribution channel, are employed to evaluate the SAPA performance of our proposed approach. The multi-tone narrowband interference from \cite{aj} was randomly added to the first $N_r$ subcarriers with better channel responses during each simulation. 2000 Monte Carlo trials were carried out to produce average performance curves. These two cases of channel responses with interference are depicted in Fig. \ref{Fig1} (a) and Fig. \ref{Fig2} (a), respectively. 

\begin{figure}[!t]
	\centering
	\subfloat[]{\includegraphics[width=1.7in]{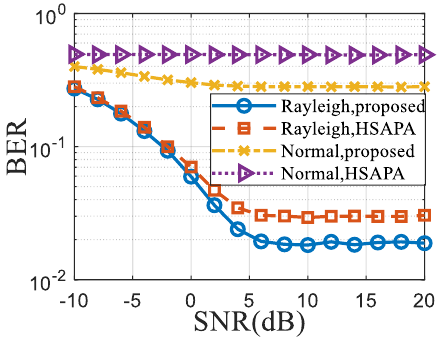}\label{fig3a}}
	\subfloat[]{\includegraphics[width=1.7in]{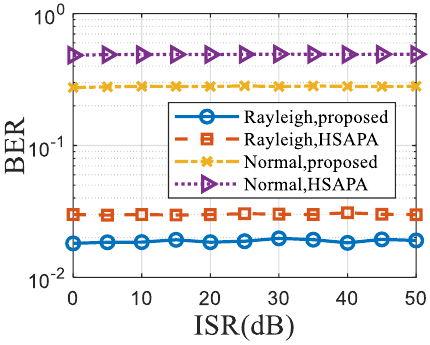}\label{fig3b}}
	\caption{Comparison of interference-resilient performance: (a) BER versus SNR, (b) BER versus ISR.}
	\label{Fig3}
\end{figure}

The traditional high-response SAPA (HSAPA) method is chosen as a baseline method for comparison, which selects the first $N_r$ subcarriers with better channel responses to transmit information \cite{ofdm1}, and the power allocation problem is the same as the proposed method. Its optimization results are shown in Fig. \ref{Fig1} (b) and Fig. \ref{Fig2} (b), where dark-blue bars represent optimized subcarrier power, and red circles indicate selected communication channels. The HSAPA method selects subcarrier channels with high responses for communication, achieving high CDR values of 58.5574 bps/Hz and 60.3964 bps/Hz for the two cases. Our proposed approach incorporates a constraint on the communication subcarrier interval, yielding results in Fig. \ref{Fig1} (c) and Fig. \ref{Fig2} (c). As can be seen, the selected communication channels are more evenly dispersed across the entire bandwidth. While the CDR values of our approach decrease to 52.3644 bps/Hz and 51.3620 bps/Hz for the above two cases, the interference-resilient performance obviously surpasses that of the HSAPA method. Fig. \ref{Fig3} (a) shows the comparison results of BER under different SNR values when the interference-to-signal ratio (ISR) is 30 dB. It can be seen that the BER performance of the proposed method is superior to HSAPA due to the reduced probability of interference. Fig. \ref{Fig3} (b) depicts the BER performance versus ISR when SNR is 10 dB, which indicates that the proposed method has robust interference-resilient ability.




\begin{figure}[!t]\vspace{-0.6cm}
	\centering
	\subfloat[]{\includegraphics[width=1.7in]{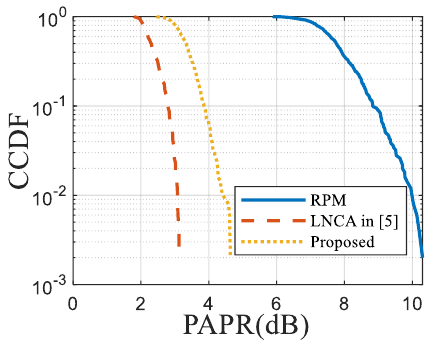}\label{fig4a}}
	\subfloat[]{\includegraphics[width=1.7in]{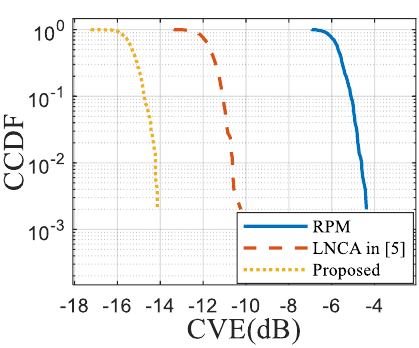}\label{fig4b}}
	\caption{Comparison of envelope performance: (a) CCDF versus PAPR, (b) CCDF versus CVE.}
	\label{fig4}
\end{figure}

Further, to analyze the envelope of the optimized waveform, we use the quadruple sampling rate so that the discrete-time envelope approximates the continuous-time envelope well \cite{papr2}. And the average results of 500 Monte Carlo trials are considered, in which the SAPA result in Fig. \ref{Fig2}(c) is used. In Fig. \ref{fig4}, the complementary cumulative distribution functions (CCDFs) of PAPR and CVE are examined by comparing the design method via $l$-norm cyclic algorithm (LNCA) in \cite{papr1} and the random phase method (RPM). Obviously, the proposed method optimizes the CVE and then indirectly reduces the PAPR, so the PAPR result is better than that of RPM, and is close to that in \cite{papr1}. Fig. \ref{fig4}(b) highlights that the CVE result of the proposed method outperforms the other two methods. Moreover, considering the lack of communication performance optimization in \cite{papr1}, the comprehensiveness of the proposed method is illustrated.

\subsection{Impact of IFPs and Parameters}\label{APA}
In this part, we set different IFPs and adaptive iterative factors under the communication channel displayed in Fig. \ref{Fig2} (a) to depict the enhancements of the ACM algorithm.

To illustrate the influence of different IFPs on the algorithm, three different cases are considered: Case 1 - $u_n^{\left( 0 \right)}=1,n = 0:{N \mathord{\left/
		{\vphantom {N {{N_r}}}} \right.
		\kern-\nulldelimiterspace} {{N_r}}}:{{\left( {N_r - 1} \right)N} \mathord{\left/
		{\vphantom {{\left( {N_r - 1} \right)N} {{N_r}}}} \right.
		\kern-\nulldelimiterspace} {{N_r}}}$; Case 2 - $u_n^{\left( 0 \right)} = 1,n = 0:\left( {L + 1} \right):\left( {L + 1} \right)\left( {{N_r} - 1} \right)$; Case 3 - $u_n^{\left( 0 \right)}=1,n = 2:{N \mathord{\left/
		{\vphantom {N {{N_r}}}} \right.
		\kern-\nulldelimiterspace} {{N_r}}}:2+{{\left( {N_r - 1} \right)N} \mathord{\left/
		{\vphantom {{\left( {N_r - 1} \right)N} {{N_r}}}} \right.
		\kern-\nulldelimiterspace} {{N_r}}}$. The average CDR and PSL results of 50 trials are summarized in TABLE \ref{tab0}. Evidently, different IFPs can obtain similar S\&C results, indicating that the proposed algorithm is effective.
\begin{table}[!t]\vspace{-0.4cm}
	\centering  
	\fontsize{8}{10}\selectfont  
	\caption{Comparison of results from different IFPs.}  
	\label{tab0}
	\begin{tabular}{|c|c|c|c|}  
		\hline
		 &Case 1 &Case 2 &Case 3\\
		\hline
		CDR(bps/Hz) &51.2764 &51.2425 &51.1818\\
		\hline
		PSL(dB) &-22.0799 &-22.0804 &-22.0812\\
		\hline
	\end{tabular} 
\end{table}

\begin{figure}[!t]
	\centerline{\includegraphics[width=3in]{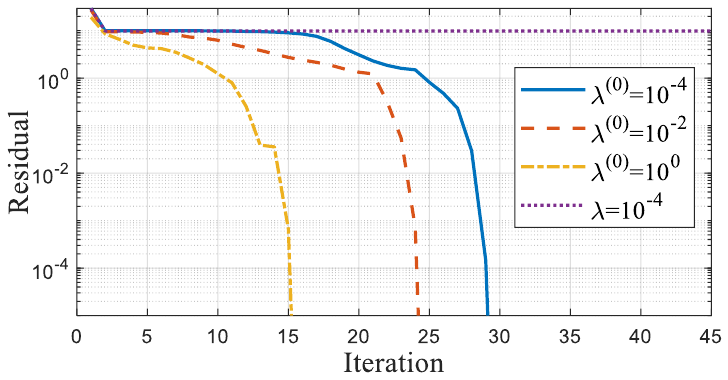}}
	\caption{Convergence of step 2 in the ACM algorithm: residual versus iteration index.}
	\label{fig5}
\end{figure}

In addition, we set ${\xi _1}=0.9$, ${\xi _2}=2$, ${\lambda ^{\left( 0 \right)}} = {10^{ - 4}},{10^{ - 2}},{10^0}$ and $\lambda  = {10^{ - 4}}$ to examine the convergence of the proposed algorithm. Fig. \ref{fig5} depicts that different initial values of $\lambda^{\left( 0 \right)}$ affect the convergence speed, but the proposed algorithm remains convergent. In contrast, if the weight factor $\lambda  = {10^{ - 4}}$ is fixed, it may not converge. Therefore, the proposed algorithm is more robust.

\section{Conclusion}\label{Conclu}
In this letter, we have devised a joint SAPA method for OFDM waveform in ISAC systems with two steps to enhance S\&C performance. Firstly, an integrated OFDM model with minimizing autocorrelation PSL and maximizing CDR under the constraints of power and communication subcarrier interval has been introduced. We have presented the ACM algorithm to solve the nonconvex optimization problem, and introduced an adaptive iterative factor to improve the convergence. Secondly, by further optimizing the phase of the complex waveform, the CVE optimization problem has been effectively solved via the LS algorithm. Finally, the effectiveness of the proposed method has been verified by numerical simulations.

\vfill

\end{document}